\documentclass[twocolumn, twocolappendix]{aastex631}

\usepackage{amsmath}
\usepackage{bm}

\newcommand\alfven{Alfv\'{e}n}

\graphicspath{{./}{Figures/}}

\begin{document}

\title{Alfv\'en Wave Conversion to Low Frequency Fast Magnetosonic Waves in Magnetar Magnetospheres}

\correspondingauthor{Dominic Bernardi}
\email{bdominic@wustl.edu}

\author[0009-0008-1825-6043]{Dominic Bernardi}

\affiliation{Physics Department and McDonnell Center for the Space Sciences, Washington University in St. Louis; MO, 63130, USA}

\author[0000-0002-0108-4774]{Yajie Yuan}
\affil{Physics Department and McDonnell Center for the Space Sciences, Washington University in St. Louis; MO, 63130, USA}

\author[0000-0002-4738-1168]{Alexander Y. Chen}
\affil{Physics Department and McDonnell Center for the Space Sciences, Washington University in St. Louis; MO, 63130, USA}

\begin{abstract}

Rapid shear motion of magnetar crust can launch Alfv\'{e}n waves into the magnetosphere. The dissipation of the Alfv\'{e}n waves has been theorized to power the X-ray bursts characteristic of magnetars. However, the process by which Alfv\'{e}n waves convert their energy to X-rays is unclear. Recent work has suggested that energetic fast magnetosonic (fast) waves can be produced as a byproduct of Alfv\'{e}n waves propagating on curved magnetic field lines; their subsequent dissipation may power X-ray bursts. In this work, we investigate the production of fast waves by performing axisymmetric force-free simulations of Alfv\'{e}n waves propagating in a dipolar magnetosphere. For Alfv\'{e}n wave trains that do not completely fill the flux tube confining them, we find a fast wave dominated by a low frequency component with a wavelength defined by the bouncing time of the Alfv\'{e}n waves. In contrast, when the wave train is long enough to completely fill the flux tube, and the Alfv\'{e}n waves overlap significantly, the energy is quickly converted into a fast wave with a higher frequency that corresponds to twice the Alfv\'{e}n wave frequency. We investigate how the energy, duration, and wavelength of the initial Alfv\'{e}n wave train affect the conversion efficiency to fast waves. For modestly energetic star quakes, we see that the fast waves that are produced will become non-linear well within the magnetosphere, and we comment on the X-ray emission that one may expect from such events.

\end{abstract}


\section{Introduction} \label{sec:intro}

Magnetars are a type of neutron star with extremely high magnetic fields. They show pulsed emission as well as a variety of transient X-ray events, such as bursts, outbursts, and giant flares \citep[see e.g.\ reviews by][]{Mereghetti_2015,2017ARA&A..55..261K}. What distinguishes them from normal pulsars is their observed luminosity, which exceeds the spin down energy budget \citep{Mereghetti_2015}. It is thought that this excess luminosity is powered by the evolution of its magnetic field \citep{1996ApJ...473..322T}.

Frequent short X-ray bursts are a hallmark of magnetar activity. These X-ray bursts range from a few milliseconds to a few seconds in duration and have peak luminosity ranging from $10^{36}$ to $10^{43}\,\rm{erg\,s^{-1}}$. \citet{1995MNRAS.275..255T} discussed these bursts in relation to dynamic activity on the surface of magnetars, arguing that the magnetic fields of these neutron stars are strong enough to crack the crust of the star and release the energy from the evolving magnetic field. The energetics of such a process are comparable to the X-ray bursts we observe \citep{1989ApJ...343..839B}. The shear motion of the magnetic foot points that follows the cracking of the crust is often referred to as a star quake.
 
Given the evolving magnetic field inherent to magnetars, the star quake model is a compelling explanation for X-ray bursts. However, the detailed mechanisms by which the energy from the star quake goes into X-ray radiation are still unclear. A typical theoretical picture is that the star quakes will fill the magnetosphere with \alfven{} waves \citep{1995MNRAS.275..255T,2020ApJ...897..173B}, which then dissipate their energy through a variety of mechanisms. \citet{2001ApJ...561..980T} argued that the \alfven{} waves will create a turbulent cascade as they interact with one another, creating smaller scale structures that eventually heat the plasma. \citet{2019ApJ...881...13L} indicated that this process is relatively slow, allowing much of the \alfven{} wave energy to seep into the crust before significant heating can occur. That energy, once transferred to the star, initiates the plastic motion of the crust \citep{2015ApJ...815...25L}. This heats the crust, and a small amount of this energy may be observed as an afterglow. \citet{2020ApJ...897..173B} argued that \alfven{} waves will become dephased as they propagate along curved magnetic field lines, which in turn develops a current that is parallel to the field lines. If the current is too large, the plasma will be insufficient to sustain such a current and a large electric field may be induced, accelerating plasma and initiating pair production. \citet{2021ApJ...908..176Y} demonstrated that \alfven{} waves propagating in a curved magnetic field geometry can convert a significant amount of their energy into fast magnetosonic (fast) waves.

Fast waves are not confined to magnetic field lines, and therefore propagate away from the star. Recently, \citet{2022arXiv221013506C} and \citet{2023ApJ...959...34B} have demonstrated that sufficiently energetic fast waves can develop electric fields comparable to the total magnetic field, leading to energetic, collisionless shocks that may power the X-ray emission that we observe. It is therefore important to understand the energy and structure of the fast waves that are indirectly produced from star quakes. In this paper, we extend the work of \citet{2021ApJ...908..176Y} by performing similar force-free simulations of \alfven{} waves launched into a dipolar magnetosphere by star quakes. In contrast with previous work, however, we consider long wave trains of \alfven{} waves on shorter magnetic field lines. In this regime, the \alfven{} waves completely fill the flux tubes they are launched in, allowing us to study how overlapping \alfven{} waves will affect the conversion to fast waves. We also study how conversion evolves after multiple reflections of the \alfven{} wave.

This paper is broken down as follows: Section \ref{sec:Methods} briefly introduces the force-free simulations performed in this paper, followed by an explanation of the numerical measurements that were made of the simulated data. Section \ref{sec:Results} discusses the specific simulations performed and their results in four subsections discussing the structure, total energy conversion, background field deformation, and long term evolution of the \alfven{} wave train. Section \ref{sec:observation} describes the implications of our results on observable features of magnetars. Section \ref{sec:discussion} summarizes the results and discusses the limitations of the simulations performed.

\section{Numerical Methods} \label{sec:Methods}

\subsection{Simulations}
We simulate the magnetosphere of a magnetar in the limit of force-free electrodynamics. Force-free is the regime in which the magnetic field energy density is significantly larger than the kinetic energy of the plasma. The force-free condition states that $\rho \mathbf{E} + \mathbf{j}\times \mathbf{B} = 0$. This, together with the ideal magnetohydrodynamic assumption that $\mathbf{E}\cdot \mathbf{B} = 0$ and Maxwell’s equations, gives rise to the force-free current \citep{1999astro.ph..2288G,2002luml.conf.....G}:
\begin{equation}
    \mathbf{j} = \frac{c}{4\pi}\mathbf{\nabla \cdot E} \frac{\mathbf{E\times B}}{B^2} + \frac{c}{4\pi}\frac{\left(\mathbf{B\cdot \nabla \times B - E\cdot \nabla \times E}\right)\mathbf{B}}{B^2}.
    \label{ffe}
\end{equation}
For the magnetosphere of a magnetar, the magnetization is extremely high, yet there is usually enough plasma to conduct the current \citep[e.g.,][]{2021ApJ...922L...7B}, so force-free is typically a good approximation. Violations of the force-free condition can occur in this context, however, when conditions reduce the background magnetic field (e.g.\ in current sheets and for energetic fast waves), or when a pair-producing gap is induced~\citep[e.g.][]{2017ApJ...844..133C}.

We use 2D axisymmetric force-free simulations of a non-rotating neutron star with a background dipole magnetic field to investigate the evolution of Alfven waves. Using our GPU-based COmuptational Force-FreE Electrodynamics code \citep[COFFEE,][]{2021ApJ...908..176Y}\footnote{https://github.com/fizban007/CoffeeGPU}, we numerically solve for the evolution of the field in spherical coordinates.

\begin{figure}[ht!]
    \centering
    \includegraphics[width=0.48\textwidth]{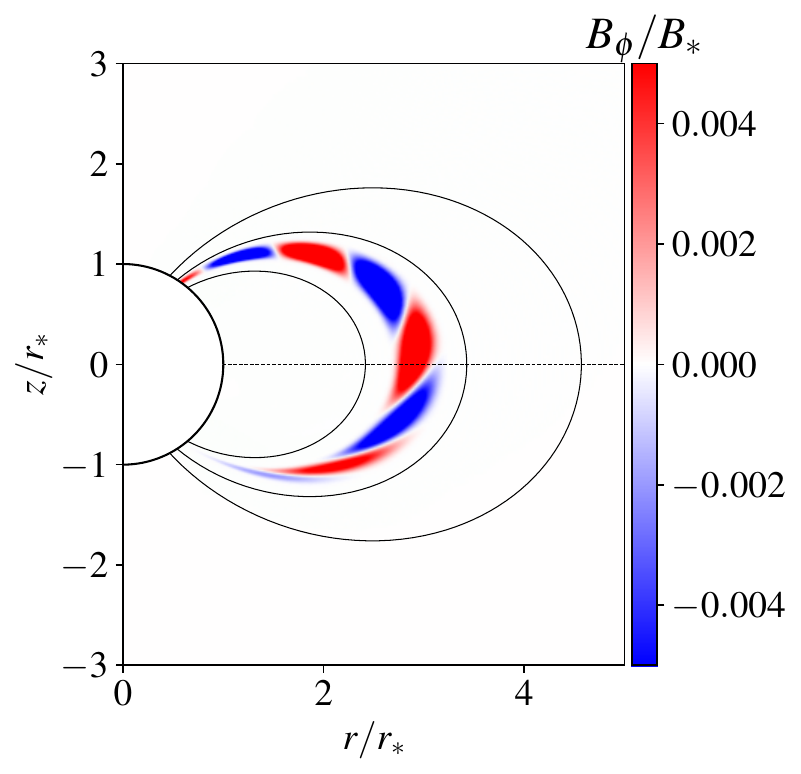}
    \caption{Alfvén waves launched by a toroidal star quake, shown at a time when the inner edge of the wave train has just reached the other end of the flux tube. The star quake perturbs the region within $\theta = 0.625 \pm 0.035$. The length of the field line at the center of the star quake is $\ell\approx6r_*$ and the duration of the star quake is $cT = \ell$. The star quake has $n = 4$ full periods and maximum amplitude $d\omega_0 = 0.092c/r_*$. In the figure, the thin black lines are poloidal magnetic field lines and the color shows $B_{\phi}$.}
    \label{fig:Alfven2D}
\end{figure}

By enforcing perfect conductivity on the stellar surface, we simulate a star quake by locally setting $\mathbf{E} = (\mathbf{r}_*\times d\omega (t)\hat{\bm{\phi}})\times \mathbf{B}$ where $d\omega(t)$ is the angular velocity of the crust. To prevent discontinuities in the field, we have a smoothing profile on $d\omega(t)$, both across $\theta$ and $t$ as follows: 
\begin{equation}
\label{eq:profile}
    d\omega(t) = d\omega_0\cos^2\left(\frac{\pi\left(\theta - \theta_c\right)}{\Delta\theta}\right)\sin^2\left(\frac{\pi t}{T}\right)\sin\left(\frac{2n\pi t}{T}\right),
\end{equation} 
for $\theta_c - (\Delta\theta/2) < \theta < \theta_c + (\Delta\theta/2)$ and $0<t<T$ where $d\omega_0$ is the maximum value of the angular velocity of the crust, $T$ is the duration of the star quake (so $cT$ is the length of the \alfven{} wave train), $n$ is the total number of wave cycles undergone during the star quake, $\theta_c$ is the center of the star quake, and $\Delta\theta$ is the angular width of the star quake. In this paper we will refer to star quakes in terms of the \alfven{} wavelength $\lambda_A = cT/n$. Figure~\ref{fig:Alfven2D} gives an example of the resultant \alfven{} wave from a simulated star quake.

\alfven{} waves launched this way propagate independently on different magnetic field lines. Due to the length difference of neighboring field lines, the wave can build up a local phase difference~\citep[see e.g.][]{2022ApJ...929...31C}, which can be seen prominently in Figure~\ref{fig:Alfven2D}. When the \alfven{} wave reaches the opposite end of the flux tube, due to our perfect conducting boundary condition at the stellar surface, the wave will reflect and eventually bounce back and forth in the flux tube. The bouncing time scale is $\ell/c$, where $\ell$ is the length of the flux tube.

When we describe our simulation results, lengths and times will be expressed relative to the stellar radius ($r_*$) and the light crossing time ($r_*/c$). We will frequently refer to the Alfvén train length ($cT$) in terms of the length of the magnetic field line ($\ell$). All fields are given in terms of the surface magnetic field strength at the equator, $B_*$. The simulations discussed typically run with a uniform grid in log($r$) with 3072 cells from $r = e^{-0.2} r_* \approx 0.8r_*$ to $r = e^{4.5} r_* \approx 90r_*$ and a uniform grid in $\theta$ with 2048 cells from $\theta=0$ to $\theta=\pi$.

\subsection{Analysis}
\label{subsec:Analysis}

The initial star quakes are entirely toroidal, launching Alfvén waves with toroidal magnetic fields. To quantify the conversion from Alfvén waves to fast waves, we need to measure the energy in each of the wave modes. The total energy in the perturbed fields is
\begin{align}\label{eq:total_energy}
    \mathcal{E}_T &= \frac{1}{8\pi}\int ((\mathbf{B}_{0}+\delta \mathbf{B})^2 + \delta \mathbf{E}^2 - \mathbf{B}_{0}^2)\, dV\nonumber\\
    &= \frac{1}{8\pi}\int (\delta \mathbf{B}^2 + \delta \mathbf{E}^2 + 2\mathbf{B}_{0}\cdot\delta \mathbf{B})\, dV,
\end{align}
where $\delta \mathbf{E}$ and $\delta \mathbf{B}$ are the electric and magnetic field of the perturbation, and $\mathbf{B}_{0}$ is the background magnetic field which is a constant magnetic dipole. For the last cross term in equation (\ref{eq:total_energy}), note that
\begin{equation}
\label{Cross-term}
    \begin{split}
        \frac{d}{dt}\int \mathbf{B}_{0}\cdot\delta \mathbf{B}\, dV &= -c\int  \mathbf{B}_{0}\cdot\left(\nabla \times \delta \mathbf{E}\right)\,dV\\
        &=c\oint \left(\mathbf{B}_{0}\times\delta \mathbf{E}\right)\cdot d\mathbf{S}=0,
    \end{split}
\end{equation}
because $\delta\mathbf{E}$ on the stellar surface is either in the poloidal direction (during the \alfven{} wave injection) or zero (after the injection has ended). Therefore, the contribution of the cross term to the energy remains zero, and the total energy is just
\begin{equation}
    \mathcal{E}_T = \frac{1}{8\pi}\int \left(\delta B^2 + \delta E^2\right) dV.
\end{equation}

In axisymmetric simulations, we can further separate the Alfvén wave energy and the fast wave energy by exploiting their polarization. \alfven{} waves have magnetic fields perpendicular to the $\mathbf{k}-\mathbf{B}$ plane, while fast waves have their electric fields perpendicular to the $\mathbf{k}-\mathbf{B}$ plane. Based on this we can write the \alfven{} wave energy as
\begin{equation}
\label{Alfven}
    \mathcal{E}_{A} = \frac{1}{8\pi}\int \left( B_\phi^2 + E_\mathrm{pol}^2\right) dV.
\end{equation}
For the fast wave energy, one might naively use $(1/8\pi)\int(\delta B_\mathrm{pol}^2 + E_{\phi}^2)dV$, but there are two complications. Firstly, fast waves may become nonlinear when they propagate to large radius, leading to $E>B$, a violation of the force-free condition. The numerical algorithm then cuts away the portion of $E$ that exceeds $B$, artificially reducing the energy in the fast waves. Secondly, as we will discuss in section \ref{subsec:Deformation}, $\delta B_\mathrm{pol}$ not only contains the magnetic field of the fast wave, but also the deformation of the background flux tube, which is a non-propagating component. To get a more accurate measurement of the fast wave energy, we instead calculate the fast wave energy flux through a spherical surface outside of the flux tube where the \alfven{} wave propagates, and well before the fast wave starts to get nonlinear. We then integrate the flux over time to get the total energy that goes into fast waves, namely,
\begin{equation}
\label{Fast}
    \mathcal{E}_{F} =\frac{c}{4\pi} \int_0^t\int \left(\delta \mathbf{B}_\mathrm{pol} \times \mathbf{E_\phi} \right)\cdot d\mathbf{S}\, dt.
\end{equation}
Meanwhile, we define a deformation energy that accounts for the significant change in poloidal magnetic field in excess of the energy carried away by fast waves
\begin{equation}
\label{eq:Def}
    \mathcal{E}_\mathrm{Def} = \frac{1}{8\pi}\int \left(\delta B_\mathrm{pol}^2 - E_{\phi}^2\right) dV,
\end{equation}
where we have used the approximation that in the fast wave, the poloidal magnetic field $\delta B_{F, \mathrm{pol}}$ and the toroidal electric field $E_{\phi}$ are equal to each other.

\begin{figure}[ht!]
    \includegraphics[width=0.48\textwidth]{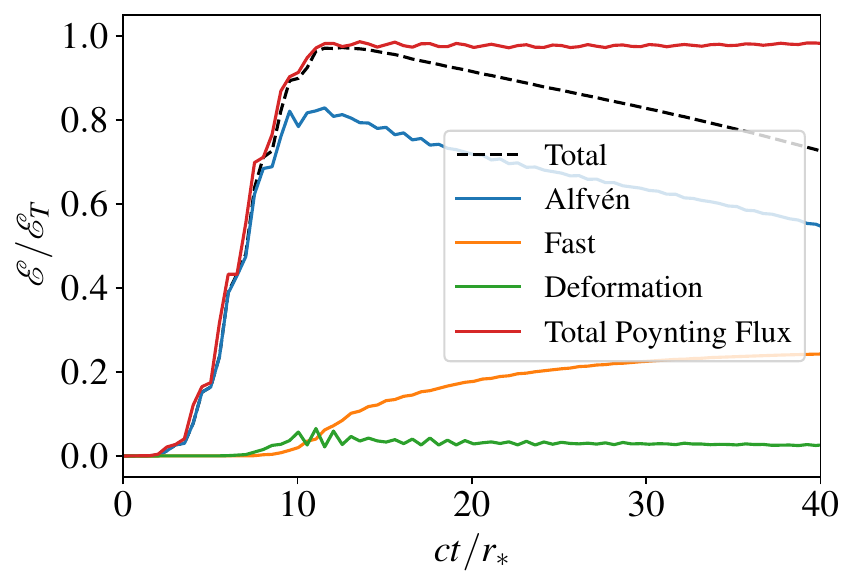}
    \caption{The evolution of energy in different wave modes relative to the maximum value of the total energy. The black dashed line, the blue line, the orange line, and the green line are the total energy, Alfvén wave energy, fast wave energy, and deformation energy, respectively. The total energy computed as a Poynting flux is also included in red. The star quake originates at $\theta = 0.8 \pm 0.05$ with a maximum angular velocity of $d\omega_0 = 0.25c/r_*$. The train length and wavelength of the injected \alfven{} wave are $cT = 4\ell$ and $\lambda_A = \ell$ respectively.}
    \label{fig:Energys}
\end{figure}

Figure \ref{fig:Energys} shows the evolution of the three different energy components as well as the total energy during a simulation. Note that since the fast wave energy is measured from the flux through a spherical surface instead of an integral over the entire domain, there is a time delay in the fast wave energy compared to other components, so at any given time step the total energy is not the sum of the three separate terms. The slow decrease in the total energy is due to numerical dissipation in the simulation, especially when the \alfven{} wave gets significantly dephased after a few reflections, and when the fast wave becomes nonlinear. Alternatively, the total energy can be calculated by integrating the energy flux going out through the stellar surface
\begin{equation}
\label{eq:TotalEnergy_flux}
    \mathcal{E}_{T} =\frac{c}{4\pi} \int_0^t\int_{r=r_*} \left(\mathbf{E} \times \mathbf{B} \right)\cdot d\mathbf{S}\, dt,
\end{equation}
as shown by the red line in Figure \ref{fig:Energys}. The maximum total energy obtained from Equation (\ref{eq:total_energy}) and Equation (\ref{eq:TotalEnergy_flux}) are consistent with each other. Throughout this paper, the total energy is computed at the end of the \alfven{} wave injection using Equation (\ref{eq:total_energy}).

\section{Results} \label{sec:Results}
\subsection{Fast Wave Structure} \label{subsec:Structure}

\begin{figure}[ht!]
    \includegraphics[width=0.49\textwidth]{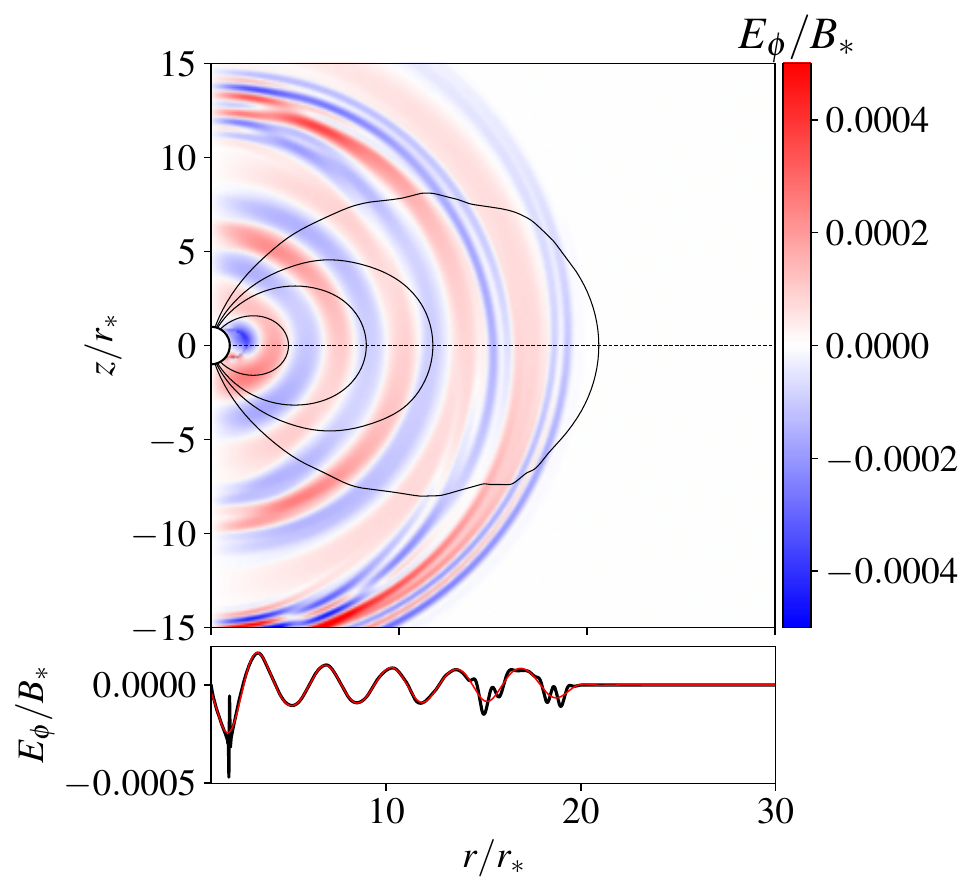}
    \caption{Top panel: 2D plot of the fast wave structure for a star quake with $\theta_c = 0.8$, $\Delta\theta = 0.1$, $d\omega_0 = 0.25c/r_*$, $cT=\ell$, and $n=2$. Bottom panel: The black line shows a cross section of $E_\phi$ along the equatorial plan obtained directly from the simulations. The red line is generated by applying a maximally flat magnitude low pass filter with a cutoff frequency of $f=2c/\ell=0.6c/r_*$ to the cross section. The spike that occurs at $r\approx 2r_*$ is within the flux tube where the fast waves are being sourced.}
    \label{fig:Ephi2D}
\end{figure}

We begin by considering star quakes with duration comparable to the bouncing time of the flux tube that confines the Alfvén waves. In section \ref{subsec:Analysis} we discussed the polarization of the \alfven{} waves and fast waves in axisymmetric force-free simulations. Given that the electric field of fast waves is entirely toroidal, we can study the structure of the fast waves by looking at $E_\phi$. We simulate a star quake in a flux tube centered at $\theta_c = 0.8$ with angular width $\Delta\theta = 0.1$. The arc length is $\ell \approx 3.3r_*$ for the magnetic field line originating at $\theta_c = 0.8$. The train length and wavelength of the \alfven{} wave are $cT = \ell$, and $\lambda_A = cT/2$. Figure \ref{fig:Ephi2D} shows $E_\phi$ after the \alfven{} wave has reflected at the stellar surface several times. The top panel is a 2D plot showing a high frequency wave front that is most dominant at the poles, and a low frequency component that becomes most apparent at later times. The bottom panel plots a 1D cross section of $E_\phi$ along the equatorial plane (plotted here in black). A maximally flat low pass filter with a cutoff frequency of $f = 2c/\ell$ is then applied to the $E_\phi$ cross section (plotted in red). The low frequency component of the fast wave is dominant even at the front of the wave and persists for several reflections. The high frequency component of the fast wave ($\lambda_{F,h}$) satisfies $\lambda_{F,h} \approx \frac{1}{2}\lambda_A$, which agrees with \cite{2021ApJ...908..176Y} and \citet{2024arXiv240406431C}. We compute the wavelength of the low frequency component of the fast wave ($\lambda_{F,l}$) and find that $\lambda_{F,l} \approx 3.5r_*$, which is roughly the length of the flux tube. The fact that $\lambda_{F,l}$ is slightly larger than the field line length $\ell\approx 3.3r_*$ at the middle of the flux tube is likely due to the \alfven{} wave being more nonlinear towards the outer portion of the flux tube.

\begin{figure}
    \centering
    \includegraphics[width=0.48\textwidth]{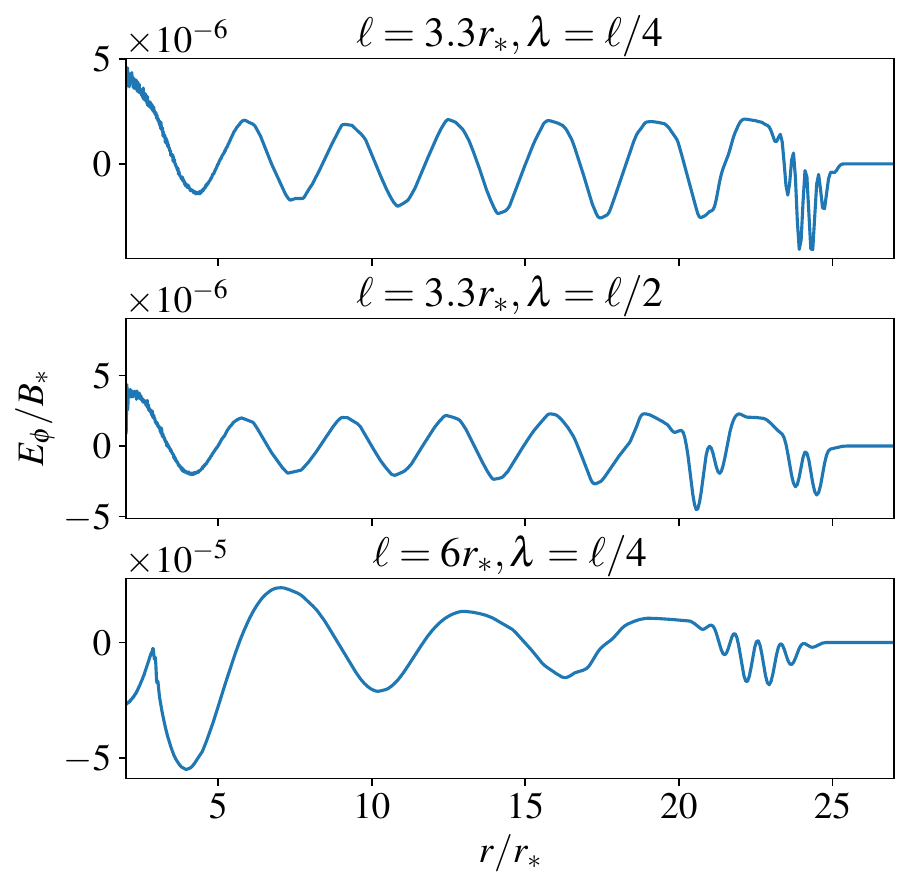}
    \caption{Cross sections of $E_\phi$ taken along $\theta = \pi/2$ at $t=25r_*/c$. The top and middle panels are produced from star quakes that perturb the region within $\theta = 0.8 \pm 0.05$ and have $d\omega_0 = 0.05c/r_*$ and $cT=\ell$. The \alfven{} wavelength of the top and middle panel are $\ell/4$ and $\ell/2$ respectively. The bottom panel is produced from a star quake launched within $\theta = 0.625 \pm 0.035$ with $d\omega_0 = 0.092c/r_*$, $cT=\ell$, and $\lambda_A = \ell/4$. The field lines that originate at $\theta_c = 0.8$ and $\theta_c = 0.625$ have lengths $\ell\approx  3.3r_*$ and $\ell\approx 6r_*$ respectively.}
    \label{fig:SlowStructure}
\end{figure}

We further demonstrate the relationship of $\lambda_F$, $\lambda_A$, and $\ell$ with a series of simulations in which we vary either $\lambda_A$ or the field line that confines the \alfven{} waves, while keeping the ratio $cT/\ell = 1$ fixed. In the top and middle panels of Figure \ref{fig:SlowStructure}, we only change the wavelength of the \alfven{} waves. For both simulations $\lambda_{F,h} \approx \lambda_A/2$, but there is little difference in the low frequency portion of the fast waves in which $\lambda_{F,l} \approx 3.5r_*$. In the bottom panel, we change $\theta_c$, increasing the length of the field line from $\ell \approx 3.3r_*$ to $\ell \approx 6r_*$. We see that the high frequency component still satisfies $\lambda_{F,h} \approx \lambda_A/2$, and the low frequency component satisfies $\lambda_{F,l} \approx \ell$.

\begin{figure}[ht!]
    \includegraphics[width=0.47\textwidth]{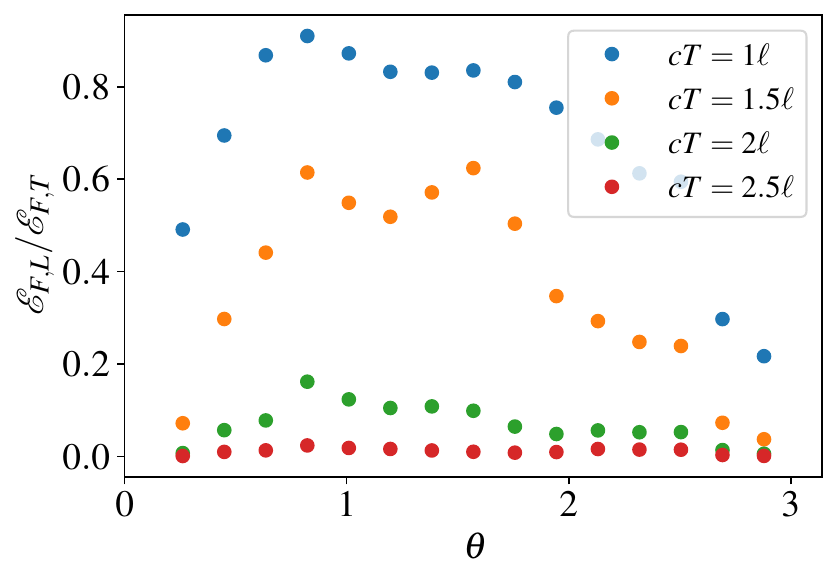}
    \caption{For several different train lengths, we compute the fraction of the energy in the low frequency fast waves. This is computed on several different cross sections of constant $\theta$. All simulations are of star quakes with $\theta_c = 0.8$, $\Delta\theta = 0.1$, $d\omega_0 = 0.25c/r_*$, and $\lambda_{A} = \ell/2$ held constant.}
    \label{fig:Ratio}
\end{figure}

We continue studying the dependence of the fast wave structure on the \alfven{} wave train lengths with a series of simulations that vary the duration $T$ of the star quakes, while keeping $\theta_c$, $\Delta\theta$, $d\omega$, and $\lambda_A$ fixed. As discussed above, Figure \ref{fig:Ephi2D} shows a 1D cross-section of the fast waves with a filtered signal that separates out the low frequency portion of the fast wave. We calculate an approximate energy in each component of the wave by integrating the square of both the original cross-section of $E_\phi$ ($\mathcal{E}_{F,T}$) and the filtered cross-section of $E_\phi$ ($\mathcal{E}_{F,L}$). We compute the ratio  $\mathcal{E}_{F,L}/\mathcal{E}_{F,T}$ at several different lines of constant $\theta$. Figure \ref{fig:Ratio} shows this ratio as a function of $\theta$ for several simulations, in which only the train length is varied. It can be seen that the low frequency component of the fast wave dominates at lower latitudes, and it is most significant for shorter \alfven{} wave trains, namely, $cT\lesssim 2\ell$. The high frequency portion of the wave is most dominant near the poles, even for train lengths $cT\sim \ell$. As the train length increases, the fast wave structure becomes dominated by high frequency waves.

To understand the transition from low frequency to high frequency dominated fast waves, we consider the mechanism that sources the two components of the wave. \cite{2024arXiv240406431C} studied \alfven{} wave propagation in a dipolar magnetosphere and found that \alfven{} waves will spontaneously convert to fast waves sourced by a toroidal current at second order in $(\delta B)/B$. This current has a component that oscillates with twice the frequency of the primary \alfven{} wave (sourcing the high frequency wave), and a constant component. Short \alfven{} wave trains that do not fill the flux tube confining them will source a low frequency fast wave as well as a high frequency wave. As the \alfven{} wave bounces in the flux tube, the relative amplitude of the wave will vary, peaking at the equatorial plane. Consequently, the constant component of $j_\phi$ will oscillate as the \alfven{} wave bounces leading to a fast wave with a frequency that depends on the bouncing time of the \alfven{} wave. \alfven{} wave trains that fill the flux tube will only produce high frequency fast waves, while the non-oscillating component of $j_\phi$ will manifest as a static inflation of the background magnetosphere. For long wave trains with $cT\gtrsim 2\ell$, the second order toroidal current $j_\phi$ mainly comes from the overlapping \alfven{} waves. We will see further evidence for this when we discuss the scaling of the fast wave production in section \ref{subsec:ConvEff}.

\subsection{Conversion Efficiency} \label{subsec:ConvEff}

\begin{figure}[ht!]
    \includegraphics[width=0.47\textwidth]{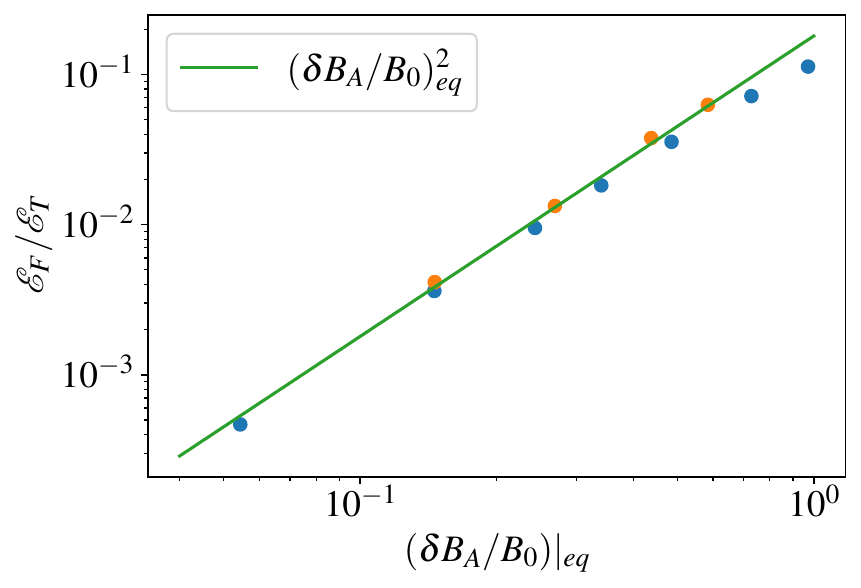}
    \caption{The conversion efficiency plotted against the relative amplitude of the Alfvén wave at the equator. The x-axis gives the theoretical relative amplitude of the center of the Alfvén wave as it passes the equatorial plane. The blue dots correspond to star quakes centered at $\theta_c = 0.8$ with $\Delta\theta = 0.1$, $cT = \ell$, and $\lambda_A = \ell/2$. The orange dots correspond to star quakes centered at $\theta_c = 0.625$ with $\Delta \theta = 0.07$, $cT = \ell$, and $\lambda_A = \ell/2$. The line is defined as $\mathcal{E}_F/\mathcal{E}_T = 0.18\cdot(\delta B_A/B_0)|_{eq}^2.$}
    \label{fig:AmpvEff}
\end{figure}

We run a series of simulations designed to explore the effects of different parameters on the conversion efficiency from \alfven{} to fast waves. We define the conversion efficiency as the ratio of the fast wave energy ($\mathcal{E}_F$) and the total energy injected during the star quake ($\mathcal{E}_T$). $\mathcal{E}_F$ is measured at the end of the simulation once further production has become insignificant. $\mathcal{E}_T$ is measured as the maximum value that the total energy attains.

First, we explore the relationship between the amplitude of the \alfven{} waves and the conversion efficiency. We perform a set of simulations on two separate flux tubes holding $\lambda_A = \ell/2$ and $cT = \ell$ constant for both sets of simulations. Varying $d\omega_0$, we calculate the conversion efficiency for each simulation as described in section \ref{subsec:Analysis}. Figure \ref{fig:AmpvEff} shows the conversion efficiency plotted against the theoretical relative amplitude of the \alfven{} waves at the equatorial plane $(\delta B_A/B_0)|_{eq}$. It is calculated from $(\delta B_A/B_0)|_{eq}=(\delta B_A/B_0)|_{*}(r_{eq}/r_*)^{3/2}$, where $r_{eq}$ is the maximum radius reached by the field line that originates from $\theta_c$ on the stellar surface. We consider $\delta B_A/B_0$ at the equatorial plane because field lines are geometrically the same when normalized by their equatorial radius. Plotting the relative amplitude here aligns the conversion efficiency trend independent of the field line. \citet{2024arXiv240406431C} argued that \alfven{} waves propagating on a dipole field will produce a second order toroidal current that depends quadratically on terms that are first order in $\delta B_A/B_0$. Specifically, they saw that this toroidal current will source a fast wave with $\delta B_F\propto (\delta B_A/B_0)^2$. The scaling in Figure~\ref{fig:AmpvEff} fits a quadratic scaling to all but the two most non-linear points, as we expect the simple scaling relation to deviate when $(\delta B_A/B_0)|_{eq}$ becomes large. It is difficult to calculate an accurate error on the fit, because the effects of numerical dissipation are difficult to quantify. For a simple scaling argument, however, we see good agreement with the quadratic dependence.

\begin{figure}[ht!]
    \includegraphics[width=0.47\textwidth]{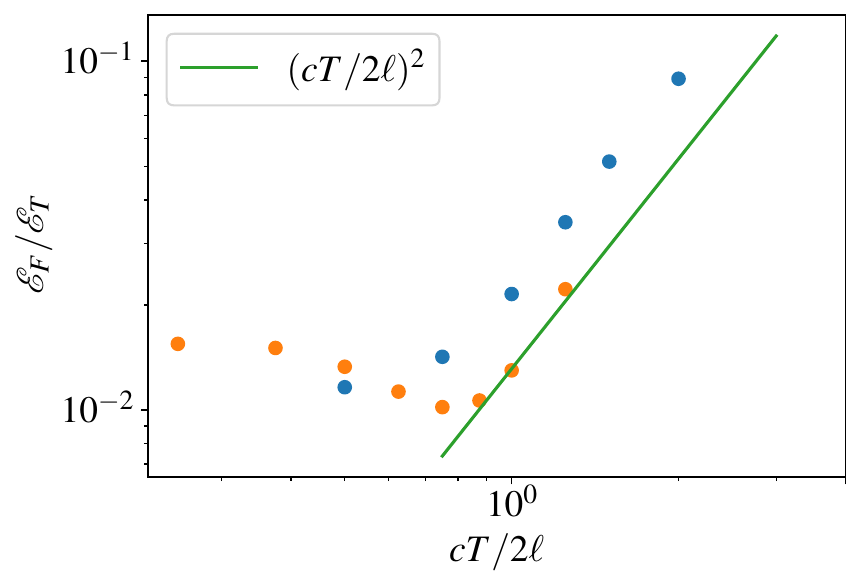}
    \caption{The conversion efficiency plotted against the train length of the \alfven{} wave launched by the star quake. The x-axis gives the train length relative to the arclength of the central field line. The orange and blue dots correspond to star quakes that perturb the region within $\theta = 0.625 \pm 0.035$, and $\theta = 0.8 \pm 0.05$ respectively. The quadratic fit is defined as $\mathcal{E}_F/\mathcal{E}_T = 0.18\cdot(\delta B_A/B_0)|_{eq}^2(cT/2\ell)^2$ where $d\omega_0$ is chosen so that $(\delta B_A/B_0)|_{eq}\approx 0.27$ for both sets of simulations. For both simulations, $\lambda_A = \ell/2$ is fixed.}
    \label{fig:TrainvEff}
\end{figure}

Next, we study how the duration of the \alfven{} wave affects the conversion efficiency to fast waves. For two different flux tubes, we perform a series of simulations in which the Alfvén wavelength is fixed and the train length is varied. For both field lines, $d\omega_0$ is chosen so $(\delta B_A/B_0)|_{eq}$ is the same. Figure \ref{fig:TrainvEff} shows dependence of the conversion efficiency on the train length relative to the arc length of the flux tube, as well as a quadratic line that corresponds to the scaling found in figure \ref{fig:AmpvEff}. The change in scaling as $cT\xrightarrow{}2\ell$ is consistent with the switch from a low frequency dominated fast wave to a high frequency dominated fast wave and indicative of a separate mechanism for producing these fast waves. Our hypothesis for the existence of the two regimes is that the low frequency fast mode is produced when the \alfven{} wave bounces back and forth along the flux tube: the pressure on the flux tube due to the \alfven{} wave varies as it moves around, leading to oscillation of the flux tube and the production of the low frequency fast waves. This will be most prominent for short wave trains. For long wave trains, the overlapping of the \alfven{} waves produces dominantly the high frequency fast mode. According to \citet{2024arXiv240406431C}, all the overlapping waves will contribute to the second order toroidal current that sources the fast wave, so we expect the fast wave amplitude to be $\delta B_F\propto cT/\ell$, where $cT/\ell$ characterizes the amount of overlap. The train length of the fast wave will be roughly proportional to $cT$ in this case (see section \ref{subsec:long_term}), so we have $\mathcal{E}_F\propto (cT)^3$, while the \alfven{} wave energy $\mathcal{E}_A\propto cT$. Therefore, $\mathcal{E}_F/\mathcal{E}_A\propto (cT)^2$ in this regime.

\begin{figure}[ht!]
    \includegraphics[width=0.47\textwidth]{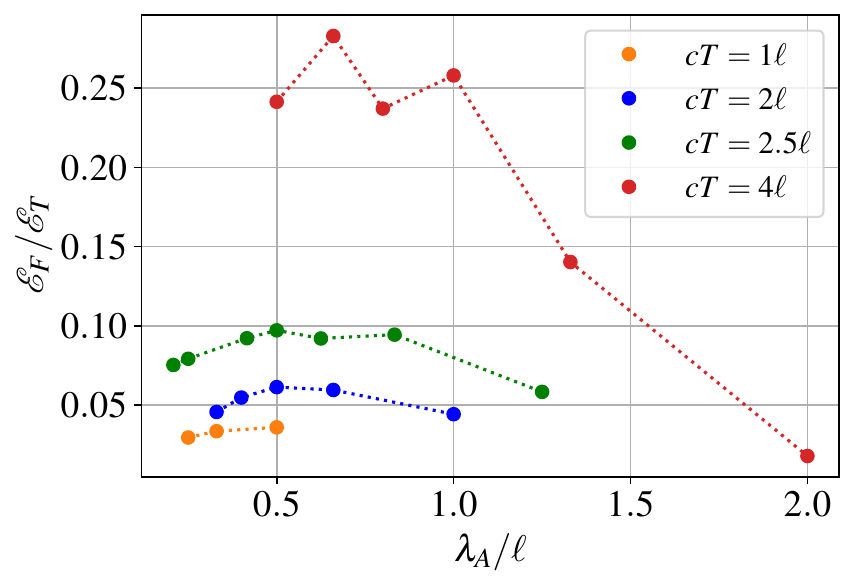}
    \caption{The conversion efficiency plotted against the wavelength of the injected \alfven{} waves. All the simulations are of star quakes with $d\omega_0 = 0.25c/r_*$ centered on the field line originating at $\theta_c = 0.8$ with an angular width of $\Delta\theta = 0.1$. The duration of the star quake is fixed for each set of simulation at $cT = 1\ell$, $cT=2\ell$, $cT = 2.5\ell$, and $cT=4\ell$.}
    \label{fig:LambdavEff}
\end{figure}

The regime switch that occurs as $cT\xrightarrow{}2\ell$ marks a transition to a high frequency dominated fast wave, the structure of which is determined by the \alfven{} wavelength. Across a series of simulations, we explore the conversion efficiency as a function of the \alfven{} wavelength in both of these regimes. We vary $\lambda_A$ for multiple different fixed train lengths. For all sets of simulations, waves are launched from a single location with fixed amplitude. The behavior of the conversion efficiencies for these simulations is plotted in Figure~\ref{fig:LambdavEff}. For $\lambda_A<\ell$, the conversion efficiency does not have a significant dependence on the wavelength. However, there is a significant decrease in conversion for $\lambda_A>\ell$. This can be understood as long wavelength \alfven{} waves beginning to approximate an adiabatic twist in the magnetosphere, a process which does not directly produce significant outgoing fast waves \citep{1986ApJ...307..205L,1995ApJ...443..810W,2013ApJ...774...92P}.

\subsection{Deformation Energy}
\label{subsec:Deformation}

When the Alfvén wavelength becomes longer than the length of the flux tube, the waves begin to behave more like adiabatic twisting of the magnetosphere. Typically, for a slowly twisted magnetosphere, the steady injection of $B_\phi$ creates a toroidal current, which will generate a poloidal magnetic 
field \citep{2009ApJ...703.1044B}. This additional poloidal field modifies the dipole background, manifesting as an inflation of the magnetosphere. We expect that the drop in conversion efficiency seen in Figure~\ref{fig:LambdavEff} at large wavelengths is due to the large wavelength beginning to approximate the slow-twist limit. We expect that in such a limit, even though there is appreciable change in the poloidal magnetic field, the change is not associated with a fast wave and does not propagate to large distances.

\begin{figure}[ht!]
    \includegraphics[width=0.49\textwidth]{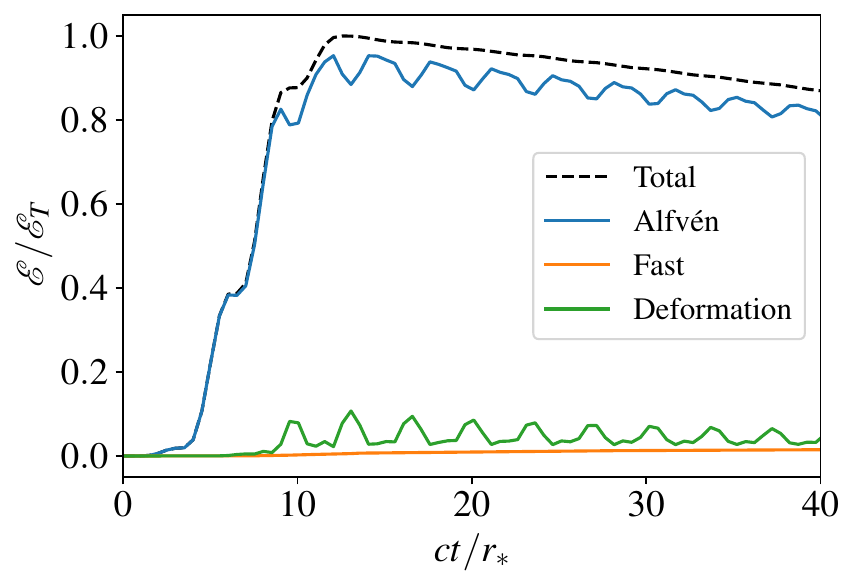}
    \caption{The evolution of different energy components relative to the maximum value of the total injected energy, for a simulation where the \alfven{} wavelength and train length are both larger than the flux tube length. The black dashed line, the blue line, the orange line, and the green line are the total energy, \alfven{} wave energy, fast wave energy, and the deformation energy, respectively. The star quake originates at $\theta = 0.8 \pm 0.05$ with a maximum amplitude of $d\omega_0 = 0.25c/r_*$, a train length $cT=4\ell$, and \alfven{} wavelength $\lambda_A = 2\ell$.}
    \label{fig:EnergyL2}
\end{figure}

\begin{figure}[ht!]
    \includegraphics[width=0.47\textwidth]{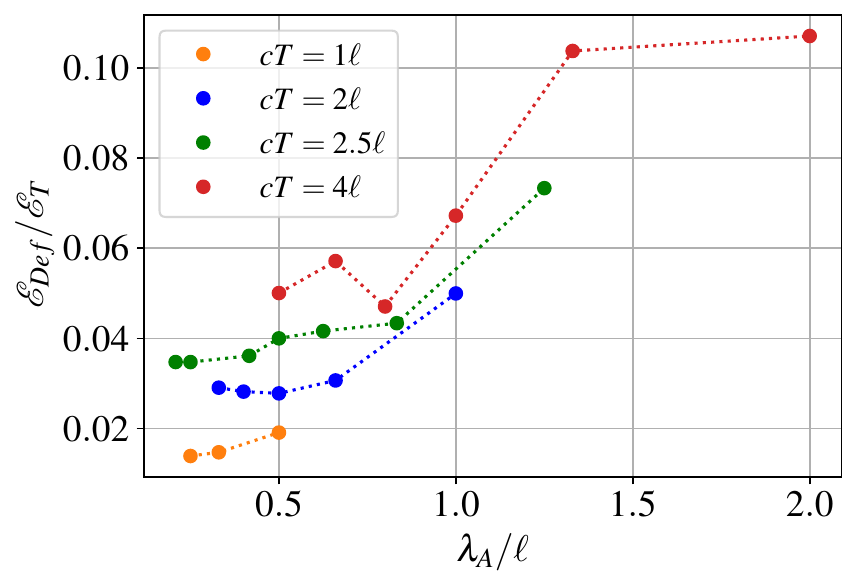}
    \caption{The deformation efficiency plotted against the wavelength of the injected \alfven{} waves. As in Figure \ref{fig:LambdavEff}, $\theta_c = 0.8$, $\Delta\theta = 0.1$, $d\omega_0 = 0.25c/r_*$, and the \alfven{} wave train length is held constant for each set of simulations at $cT = 1\ell$, $cT=2\ell$, $cT = 2.5\ell$, and $cT=4\ell$.}
    \label{fig:DefEff}
\end{figure}

To quantify this effect, we defined the deformation energy in Section~\ref{sec:Methods} as Equation~\eqref{eq:Def}. The deformation energy accounts for a change in poloidal field in excess of the fast wave energy. Figure \ref{fig:EnergyL2} shows the deformation energy and \alfven{} wave energy for a simulation with $\lambda_A = 2\ell$. In contrast with Figure~\ref{fig:Energys}, we see a significant suppression of fast wave production, and instead energy is converted to $B_\mathrm{pol}$, which oscillates out of phase with the Alfvén wave energy. We define the deformation efficiency as the ratio of the maximum deformation energy to the total injected energy. This is plotted as a function of wavelength in Figure~\ref{fig:DefEff} for the same simulations shown in Figure~\ref{fig:LambdavEff}. We see the deformation energy increases quickly after $\lambda_A = \ell$, as we expect.

\subsection{Long term evolution}
\label{subsec:long_term}
We now discuss the evolution of fast wave production after the \alfven{} wave has had several reflections on the stellar surface. The low frequency fast wave and the high frequency fast wave behave differently in their evolution. On one hand, once the fast waves are far from the flux tube, their expansion becomes approximately spherical. At this point, the amplitude of the fast waves will fall off as $1/r$. In Figure~\ref{fig:SlowStructure}, the amplitude of the radial cross sections of the low frequency component remains roughly constant. From these observations we deduce that the production of low frequency fast waves decreases with time; the amplitude at the tip of the flux tube where the wave is first generated depends on time approximately as $1/t$.

\begin{figure}[ht!]
    \includegraphics[width=0.47\textwidth]{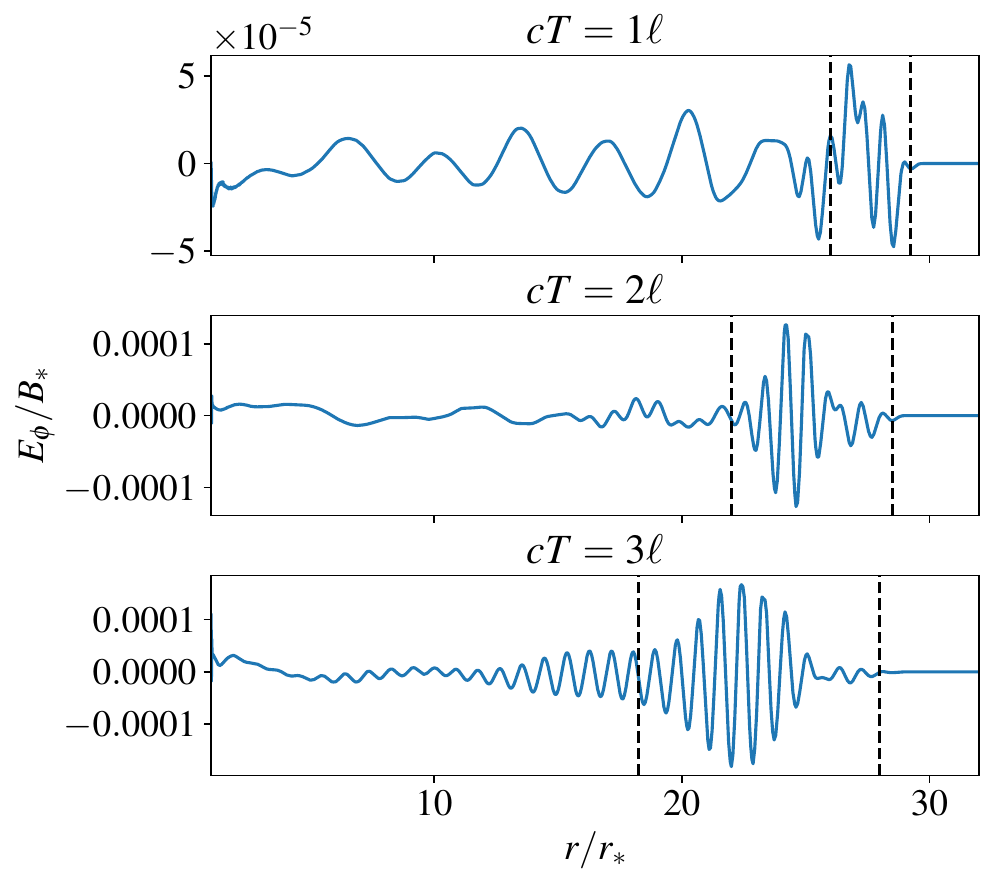}
    \caption{Cross sections of $E_\phi$ taken along $\theta = 3\pi/4$ at time $t=30r_*/c$ for simulations with different train lengths. The star quakes are initialized with $\theta_c = 0.8$, $\Delta\theta = 0.1$, $d\omega_0 = 0.138c/r_*$, and $\lambda_A = \ell/2$. The train length for each simulation is $cT = 1\ell$, $cT = 2\ell$, and $cT = 3\ell$ for the top, middle, and bottom panel respectively. The two black dotted lines in each graph are separated by one train length of the star quake.}
    \label{fig:TrainStructure}
\end{figure}

On the other hand, for the high frequency fast wave, most of the energy seems to be concentrated at the beginning. In Figure \ref{fig:TrainStructure}, we plot radial cross sections of simulations with different \alfven{} wave train lengths, while $\theta_c$, $\Delta\theta$, $\lambda_A$, and $d\omega_0$ are fixed. We see that even for short train lengths, the high frequency component of the fast waves is suppressed after about one train length. 
This suppression is likely due to the dephasing of the \alfven{} waves---waves propagating on different field lines have different bouncing times, so the wave front becomes increasingly oblique \citep{2020ApJ...897..173B, 2022ApJ...929...31C}. \cite{2021ApJ...908..176Y} found that dephasing of the \alfven{} wave does reduce the production of the high frequency fast wave. Although the de-phasing time is the same for all of the star quakes in Figure~\ref{fig:TrainStructure}, the star quake will continuously produce coherent \alfven{} waves for the duration of the star quake, thus increasing the amount of time before the wave is completely de-phased. The duration of the high frequency fast waves turns out to be approximately $T$.

\begin{figure}[ht!]
    \includegraphics[width=0.47\textwidth]{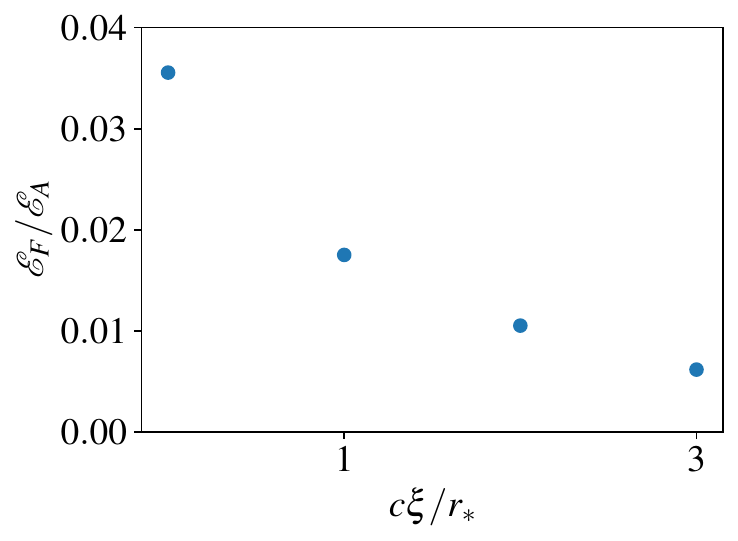}
    \caption{The dependence of the conversion efficiency on the initial shear of the wave. The star quakes are all initialized with $\theta_c = 0.8$, $\Delta \theta = 0.1$, $d\omega_0 = 0.25c/r_*$, $cT=\ell$, and $\lambda_A = \ell/2$.}
    \label{fig:Shear}
\end{figure}

To further demonstrate the role de-phasing has on the conversion, we initialize the simulations with \alfven{} waves that are already de-phased. We do this by adding a time delay ($\xi$) that varies linearly with the launch angle of the star quakes. This changes equation \ref{eq:profile} by letting $t\xrightarrow{}\tau$ where $\tau = t - \xi(\theta_2 - \theta)/(\theta_2 - \theta_1)$ and $\xi$ is the phase delay across the perturbed region. The perturbation is applied for $0<\tau<T$. For star quakes launched from fixed foot points, with fixed amplitude, wavelength, and duration, we increase $\xi$, and measure the efficiency of fast wave production. Figure \ref{fig:Shear} shows a significant reduction in conversion for highly sheared \alfven{} waves. The reduction happens for both the high frequency component and the low frequency component.

Since the dephasing time depends on the angular width $\Delta\theta$ of the star quake, $\Delta\theta$ will affect the conversion efficiency. We include a detailed study of the effect of $\Delta\theta$ on the conversion efficiency in Appendix \ref{Appendix:angular}. The main result is that the conversion efficiency for the low frequency fast waves (short \alfven{} wave train regime) is not significantly affected when $\Delta\theta$ changes; on the other hand, the high frequency fast wave production quickly drops when $\Delta\theta$ is large enough such that the \alfven{} wave inner boundary and outer boundary get out of phase with each other before one reflection, namely, the difference of the field line length at the outer and inner edge of the flux tube $\Delta\ell$ becomes larger than the \alfven{} wave length $\lambda_A$. Since $\Delta\ell\sim 3\Delta r_{eq}\sim (6r_*\cos\theta_c/\sin^3\theta_c)\Delta\theta$, this means that the high frequency fast wave production drops when 
\begin{equation}
    \label{eq:angular_limit}
    \Delta\theta\gtrsim \lambda_A\sin^3\theta_c/(6r_*\cos\theta_c).
\end{equation}

\section{Relevance to Observations}
\label{sec:observation}

Based on the results above, we can write down a crude scaling relation for the efficiency of fast wave production 
\begin{align}\label{eq:conversion_efficiency}
    \frac{\mathcal{E}_F}{\mathcal{E}_A}\approx
    \begin{cases}
    \displaystyle 0.2\left(\left.\frac{\delta B_A}{B_0}\right|_{eq}\right)^2\left(\frac{c T}{2\ell}\right)^2f\left(\theta_c, \Delta\theta, \lambda_A\right), & cT>2\ell,\\
    \displaystyle 0.2\left(\left.\frac{\delta B_A}{B_0}\right|_{eq}\right)^2, & cT<2\ell,
    \end{cases}
\end{align}
where $\delta B_A$ is the amplitude of the \alfven{} wave magnetic field, $B_0$ is the background magnetic field, $T$ is the duration of the \alfven{} wave, and $\ell$ is the length of the flux tube where the \alfven{} wave propagates. Here we include a factor $f\left(\theta_c, \Delta\theta, \lambda\right)$ to account for the effects of the angular width on the conversion efficiency. Note that this is an upper bound on the conversion efficiency for realistic star quakes, especially when a more general wave packet is considered. The low frequency fast wave is affected less significantly by dephasing and consequently has little dependence on the angular width. Another remark is that the scaling relation (\ref{eq:conversion_efficiency}) is only valid for short enough \alfven{} wavelength, $\lambda_A\lesssim\ell$. For longer wavelengths, we approach the slow twist regime, and the conversion to fast waves also drops significantly.

Now we will apply these scaling relations to the star quakes believed to be responsible for the short magnetar bursts. For a simple estimation, we will associate the duration of the burst with the duration of the star quake, and the energy radiated in X-rays would be a fraction of the energy released by the star quake. 
Suppose the \alfven{} wave launched by the star quake has a typical energy of $\mathcal{E}_A\sim 10^{40}\,{\rm erg}$, with a duration $T\sim100\,{\rm ms}$, then the initial relative amplitude of the \alfven{} wave at the stellar surface is
\begin{align}\label{eq:alfven_initial_rel_amp}
    \left.\frac{\delta B_A}{B_0}\right|_*&\approx \frac{1}{B_*r_*}\sqrt{\frac{4\mathcal{E}_A}{cT \delta \theta}}\nonumber\\
    &\approx 1\times 10^{-4}\mathcal{E}_{A,40}^{1/2}T_{-1}^{-1/2}\delta\theta_{-1}^{-1/2}B_{*,14}^{-1},
\end{align}
where $\mathcal{E}_{A,40}=\mathcal{E}_A/(10^{40}\,\rm{erg})$, $T_{-1}=T/(10^{-1}\,\rm{s})$, and we assumed that the quake region has an angular width of $\delta\theta=10^{-1}\delta\theta_{-1}$, the surface magnetic field is $B_*=10^{14}B_{*,14}\,\rm{G}$, and the stellar radius is $r_*\sim10^6\,\rm{cm}$.
Since the relative amplitude grows with radius as
\begin{equation}
    \frac{\delta B_A}{B_0}\propto \left(\frac{r}{r_*}\right)^{3/2},
\end{equation}
the wave will become nonlinear, $\delta B_A/B_0\sim1$, at a radius
\begin{equation}
    r_{A,nl}\sim 4.2\times10^2r_* \mathcal{E}_{A,40}^{-1/3}T_{-1}^{1/3}\delta\theta_{-1}^{1/3}B_{*,14}^{2/3}.
\end{equation}
Meanwhile, if the \alfven{} wave has sufficient energy compared to the background magnetospheric energy, $\mathcal{E}_A\gtrsim B^2r^3$, then the wave can break out from the magnetosphere. The ejection radius is 
\begin{equation}
    r_{A,ej}\sim 10^2r_*\mathcal{E}_{A,40}^{-1/3}B_{*,14}^{2/3}
\end{equation}
but keep in mind that this is a very rough estimation. An \alfven{} wave needs to become nonlinear in order to eject, so $r_{A,ej}\gtrsim r_{A,nl}$.

We will be considering \alfven{} waves propagating within the region $r<r_{A,nl}$. As a comparison, the light cylinder radius of the magnetar is at $r_{LC}\sim4.8\times10^{9}P_0\,{\rm cm}\sim4.8\times10^{3}P_0r_*$, for a spin period of $P_0$ second. Therefore, slowly spinning magnetars with sufficiently energetic quakes will have $r_{A,nl}$ well within the light cylinder. The wave train length is $cT\sim 3\times10^9T_{-1}\,{\rm cm}\sim 3\times10^3T_{-1}r_*$. For a dipole flux tube with maximum radius $r_m$, the length of the flux tube is $\ell\sim 3r_m$. We can see that for quakes with short enough duration (e.g., $T\lesssim70\, {\rm ms}$ at $\mathcal{E}_A\sim 10^{40}\,{\rm erg}$), the wave train length $cT$ can be shorter than $2\ell$ when the \alfven{} wave propagates on a field line with maximum radius $r_m\sim r_{A,nl}$; otherwise the train length is usually larger than twice the flux tube length. Meanwhile, during the star quake, \alfven{} waves with frequency $\omega_A\gtrsim 10^4$ Hz can be produced and transmitted into the magnetosphere \citep{2020ApJ...897..173B}, so the wavelength of the \alfven{} waves should be $\lambda_A=2\pi c/\omega_A\lesssim2\times10^7\,{\rm cm}\sim 20r_*$.

For a localized star quake, the oscillation of the crust starts locally, then propagates to the whole star \citep{2020ApJ...897..173B}. Correspondingly, the \alfven{} waves are initially launched from the quake region, but after the crustal elastic waves have propagated, the launching region of the \alfven{} waves extends to the full star. So in reality, we need to consider \alfven{} waves propagating on a wide range of field lines starting from different latitudes. Imagine the same \alfven{} wave with initial amplitude $(\delta B_A/B_0)_*$ is launched from different polar angle $\theta_*$. The field line with foot point at $\theta_*$ will reach a maximum radius $r_m=r_*/\sin^2\theta_*$. If the wave is launched close to the equator, the field lines are short, and the train length is larger than twice the flux tube length, $cT>2\ell$. The first regime in equation (\ref{eq:conversion_efficiency}) applies. Since 
\begin{equation}
    \left.\frac{\delta B_A}{B_0}\right|_{eq}=\left.\frac{\delta B_A}{B_0}\right|_{*}\left(\frac{r_m}{r_*}\right)^{3/2}=\left.\frac{\delta B_A}{B_0}\right|_{*}\frac{1}{\sin^3\theta_*},
\end{equation}
and $\ell\approx 3r_m$, we get the conversion efficiency as a function of $\theta_*$ as
\begin{equation}\label{eq:efficiency_long}
    \frac{\mathcal{E}_F}{\mathcal{E}_A}\approx 0.2\left(\left.\frac{\delta B_A}{B_0}\right|_{*}\right)^2\left(\frac{c T}{6 r_*}\right)^2\frac{1}{\sin^2\theta_*}f.
\end{equation}
The conversion efficiency increases as $\sin\theta_*$ decreases. Note that $f$ starts to significantly drop below 1 when condition (\ref{eq:angular_limit}) is satisfied so that dephasing becomes significant. For $\lambda_A\sim 20r_*$ and $\Delta\theta\sim0.1$, this would mean $\theta_*\lesssim 0.3$ or $r_m\gtrsim11r_*$. So starting from near the equator, as the \alfven{} waves propagate on increasingly longer field lines, more energy goes into fast waves, but this starts to drop when dephasing becomes fast at small enough $\theta_*$.

\begin{figure}
    \centering
    \includegraphics[width=\columnwidth]{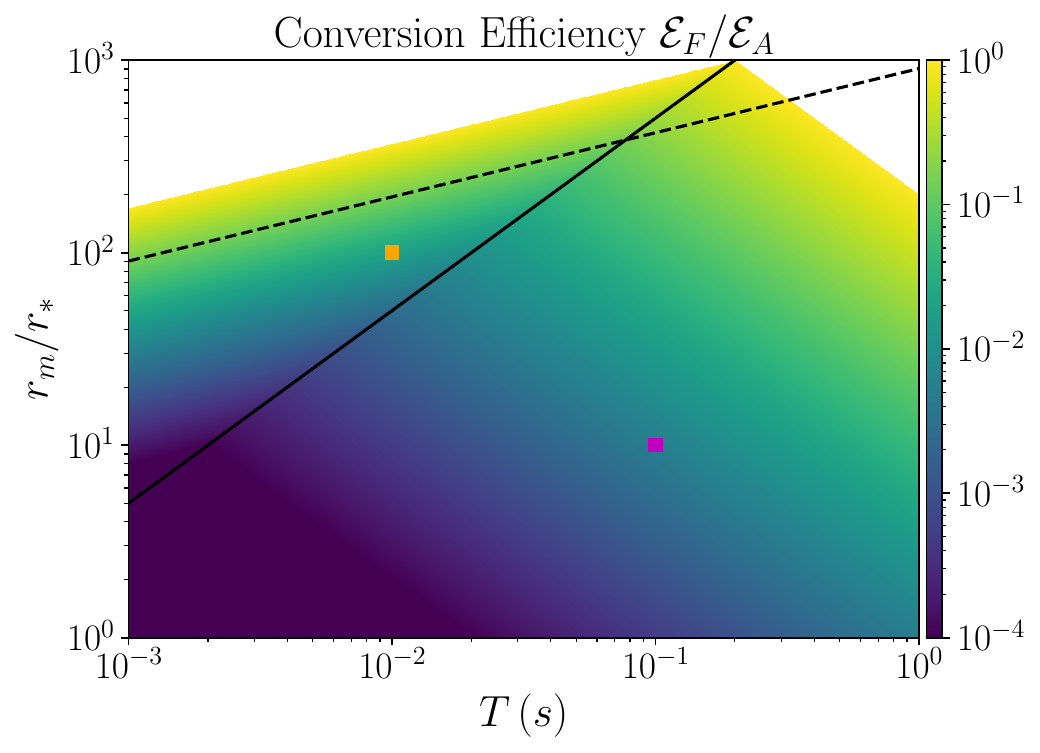}
    \caption{The parameter space for different regimes of fast wave production from \alfven{} waves. Color shows the conversion efficiency $\mathcal{E}_F/\mathcal{E}_A$. The white region marks where our scaling predicts $\mathcal{E}_F/\mathcal{E}_A > 1$ and other nonlinear phenomena should take over. Here we consider an \alfven{} wave with a fixed energy $\mathcal{E}_A\sim 10^{40}$ erg and a fixed angular width $\delta\theta\sim 0.1$ at launch, but the following two parameters can vary: duration $T$ of the star quake, and the equatorial extent $r_m$ of the field line on which the \alfven{} wave propagates. The region below the dashed black line corresponds to $r_m<r_{A,nl}$---this is the region of interest. The solid black line corresponds to $c T\sim2\ell\sim 6r_m$, which separates the short wave train regime (above) and long wave train regime (below). When $\mathcal{E}_A$ increases, the dashed line moves downwards, while the solid line remains unchanged. Magenta and orange squares correspond to the two examples we consider in the text.}
    \label{fig:parameter_space}
\end{figure}

Further decreasing $\theta_*$ leads to a different scaling relation: when $\sin\theta_*\lesssim\sqrt{6 r_*/(cT)}$,  the train length becomes smaller than twice the flux tube length, $cT\lesssim 2\ell$. In this regime, the conversion efficiency is
\begin{equation}\label{eq:efficiency_short}
    \frac{\mathcal{E}_F}{\mathcal{E}_A}\approx 0.2\left(\left.\frac{\delta B_A}{B_0}\right|_{*}\right)^2\frac{1}{\sin^6\theta_*},
\end{equation}
which increases much faster as $\sin\theta_*$ decreases.
This regime can be reached by an \alfven{} wave with a duration $T\lesssim70$ ms at a fixed energy $\mathcal{E}_A\sim 10^{40}\,{\rm erg}$, as the wave propagates on a long field line such that $cT\lesssim 2\ell$ but the wave is not yet nonlinear. The fast waves produced would have very long wavelengths, on the order of the flux tube length, $\lambda_F\sim \ell$. We find that the maximum conversion efficiency would be on the order of 20\%, which happens when the \alfven{} wave propagates on the field line with $r_m\sim r_{A,nl}$, namely, the \alfven{} wave is just about to become nonlinear. On the other hand, if the \alfven{} wave has a duration $T>70$ ms, then $cT>2\ell$ is always satisfied on all flux tubes with $r_m\lesssim r_{A,nl}$. The fast waves generated in this regime will be primarily the high frequency mode, with $\lambda_F\sim\lambda_A/2$. Figure \ref{fig:parameter_space} shows the conversion efficiency computed using Equations~\eqref{eq:efficiency_long} and \eqref{eq:efficiency_short} over a range of parameters, given a fixed energy $\mathcal{E}_A$ in the \alfven{} waves. 

With the scaling of the conversion efficiency, we can also estimate the luminosity of the fast waves. Our simulations suggest that for high frequency fast mode, most of the energy is concentrated at the front of the wave train within a segment of length $\sim cT$. Therefore, in the regime of $cT>2\ell$, we have
\begin{equation}
    L_F\sim \frac{\mathcal{E}_F}{T}.
\end{equation}
For low frequency fast mode, on the other hand, the duration is much longer. The amplitude of the generated fast wave at the starting point roughly goes as $1/t$, so we have $\mathcal{E}_F=\int_{t_0}^{\infty}L_{F0}t_0^2/t^2\,dt=L_{F0}t_0$, where $L_{F0}$ is the maximum luminosity of the fast wave at the beginning time point $t_0$, and $t_0\sim \ell/c$. For this regime, we get a rough estimation of the peak luminosity of the fast wave as 
\begin{equation}
    L_F\sim\frac{\mathcal{E}_Fc}{\ell}.
\end{equation}
The amplitude of the fast wave, when it is first generated, at a radius $r\sim r_m\sim \ell/3$, is roughly
\begin{equation}
    \delta B_F|_{r=r_m}\sim\sqrt{\frac{2L_F}{r_m^2c}}. 
\end{equation}
As the fast wave propagates in the closed zone of the magnetosphere, its amplitude relative to the dipolar background magnetic field grows as $\delta B_F/B_0\propto r^2$, so the wave can become nonlinear, namely, $\delta B_F/B_0\sim 1/2$ \citep{2023ApJ...959...34B,2022arXiv221013506C}, at a radius $r_{F,nl}\approx (2\delta B_F/B_0)_{r=r_m}^{-1/2}r_m$. In what follows, we look at some examples in the two different regimes.

In the long wave train regime, using equation (\ref{eq:efficiency_long}), we get 
\begin{equation}
    \delta B_F|_{r=r_m}\sim 5.8\times 10^7\sin\theta_* \mathcal{E}_{A,40}\delta\theta_{-1}^{-1/2}B_{*,14}^{-1} f^{1/2}\ {\rm G}.
\end{equation}
For an \alfven{} wave with an energy of $10^{40}$ erg, duration $T\sim 0.1$ s, and launched from a polar angle $\theta_*$ such that $\sin\theta_*=0.32$ on the stellar surface, we have $r_m=10 r_*$, $cT/\ell\sim 100$---this corresponds to the magenta point in Figure \ref{fig:parameter_space}, located well within the long wave train regime, and dephasing is still insignificant. Since $(\delta B_A/B_0)_{eq}\sim 0.1$, we get  $\mathcal{E}_F/\mathcal{E}_A\approx 0.005$, $\delta B_F|_{r=r_m}\sim 1.8\times 10^7\,\mathrm{G}$, and the fast wave gets nonlinear at a radius $r_{F,nl}\sim5.2\times10^2r_*$, well within the light cylinder if the period of the magnetar is $P\sim 1$ s. 

In the short wave train regime, using equation (\ref{eq:efficiency_short}), we obtain
\begin{equation}
    \delta B_F|_{r=r_m}\sim 3.7\times 10^6 \mathcal{E}_{A,40}T_{-1}^{-1/2}\delta\theta_{-1}^{-1/2}B_{*,14}^{-1}\ {\rm G}.
\end{equation}
For an \alfven{} wave with an energy of $10^{40}$ erg, duration $T\sim 10$ ms, propagating on a field line with length $\ell\sim cT$, or $r_m\sim 10^2r_*<r_{A,nl}$, we have $(\delta B_A/B_0)_{eq}\sim 0.32$, $\mathcal{E}_F/\mathcal{E}_A\approx 0.02$, corresponding to the orange point in Figure \ref{fig:parameter_space}. The peak magnetic field in the fast wave is reached at the beginning, with $\delta B_F|_{r=r_m}\sim 1.1\times 10^7~\mathrm{G}$, and this portion of the fast wave gets nonlinear at a radius $r_{F,nl}\sim2.1\times10^2r_*$. 
In general, we find that for a large parameter space, the converted fast wave will become nonlinear within the light cylinder of the magnetar. Note that we used a modest \alfven{} wave energy of $10^{40}$ erg in the above estimations. For more energetic \alfven{} waves as required by bright X-ray bursts, we should expect much more efficient conversion to fast waves, because the dependence on \alfven{} wave amplitude is quadratic in Equation (\ref{eq:conversion_efficiency}).

The nonlinear evolution of fast waves in the magnetosphere can lead to efficient dissipation of the wave energy. However, the physics of this dissipation process cannot be captured in our force-free simulations. \citet{2022arXiv221013506C} and \citet{2023ApJ...959...34B} have shown that in a plasma filled magnetosphere, when the fast waves become nonlinear, they will launch shocks at every wavelength, which will dissipate at least half of the energy in the fast waves; particles heated by the shocks will radiate efficiently in X-rays. \citet{2022arXiv221013506C} also showed that when the plasma supply in the magnetosphere is low, the nonlinear fast waves can have regions with $E>B$ which accelerate particles to high energies and they will then quickly radiate and cool. In both regimes, the dissipated fast wave energy will go into X-ray emission, contributing to the emission observed during sufficiently energetic X-ray bursts. Note that we did not take into account the effect of magnetar rotation in this study; rotation may lead to more efficient production of fast waves, especially when the \alfven{} waves propagate on flux tubes far away from the star/close to the light cylinder \citep{2024arXiv240406431C}.

\section{Conclusion and Discussion}
\label{sec:discussion}

We have detailed the results of a series of 2D force-free simulations of \alfven{} waves propagating in a static dipolar magnetosphere, focusing on the production of fast waves as the \alfven{} waves bounce back and forth along a curved flux tube. The fast waves produced in this process can be split into two regimes: a low frequency dominated fast wave whose wavelength corresponds to the arc length of the flux tube, and a high frequency fast wave whose wavelength corresponds to half of the \alfven{} wavelength. The transition from low frequency dominated fast waves to high frequency dominated fast waves occurs when the train length of the generative \alfven{} wave train becomes longer than about twice the arc length of the confining flux tube. We quantified how the efficiency of fast wave production scales with the amplitude, duration and wavelength of the initial \alfven{} wave (Equation \ref{eq:conversion_efficiency}). We find that for sufficiently energetic quakes, both regimes could see fast waves becoming nonlinear within the light cylinder of the magnetar, which could lead to strong dissipation and X-ray emission.

The force-free simulations we consider here have demonstrated the existence of a low frequency fast wave as well as emphasized the non-trivial effects of significantly overlapping fast waves. However, the approximations made in this simulation do not give a complete picture of star quakes. Axisymmetric force-free simulations of a non-rotating dipolar magnetosphere is a simple approximation for short field lines within the closed zone of a magnetar magnetosphere. For field lines that approach the light cylinder, effects of rotation may significantly alter the conversion to fast waves. \citet{2021ApJ...908..176Y} and \citet{2024arXiv240406431C} demonstrated that the high frequency fast wave will instead have a frequency equal to the frequency of the \alfven{} wave. It is unclear, however, how the low frequency fast wave will be effected by the inclusion of rotation. Relaxing the axisymmetric approximation may also effect the conversion to fast modes because more wave modes are allowed to propagate, however the modes would no longer be separated by their polarization. This makes the problem of \alfven{} wave conversion to fast waves difficult to study, especially in inclined magnetospheres, but may also allow for more general wave interactions that may significantly change the production of fast waves.

We also emphasize that \alfven{} wave conversion to fast waves is only one route by which their energy is dissipated. Section \ref{sec:intro} discussed many of the other mechanisms by which \alfven{} waves can transfer their energy. Some of these mechanisms, such as dissipation through the crust of the star, will contribute while fast wave production is occurring. Consequently, these simulations only demonstrate a limited picture of what true star quakes would look like. Many of the effects are not well captured by force-free simulations. For a complete picture of both the dissipation of \alfven{} waves and the ultimate fate of the fast waves, it is necessary to carry out the study in a kinetic framework.

\begin{acknowledgments}

We thank Yuanhong Qu for helpful discussion. We also thank the anonymous referee for their insight. DB acknowledges training provide by the NSF/APS-DPP GPAP summer school. AC and YY acknowledge support from NSF grants DMS-2235457 and AST-2308111. This work was also facilitated by Multimessenger Plasma Physics Center (MPPC), NSF grant PHY-2206608, and by a grant from the Simons Foundation (MP-SCMPS-00001470) to YY. This research used resources of the Oak
Ridge Leadership Computing Facility at the Oak Ridge National Laboratory,
which is supported by the Office of Science of the U.S. Department of Energy
under Contract No. DE-AC05-00OR22725.

\end{acknowledgments}

\vspace{5mm}

\appendix

\section{Angular width dependence}
\label{Appendix:angular}

\begin{figure}[ht]
    \includegraphics[width=0.47\textwidth]{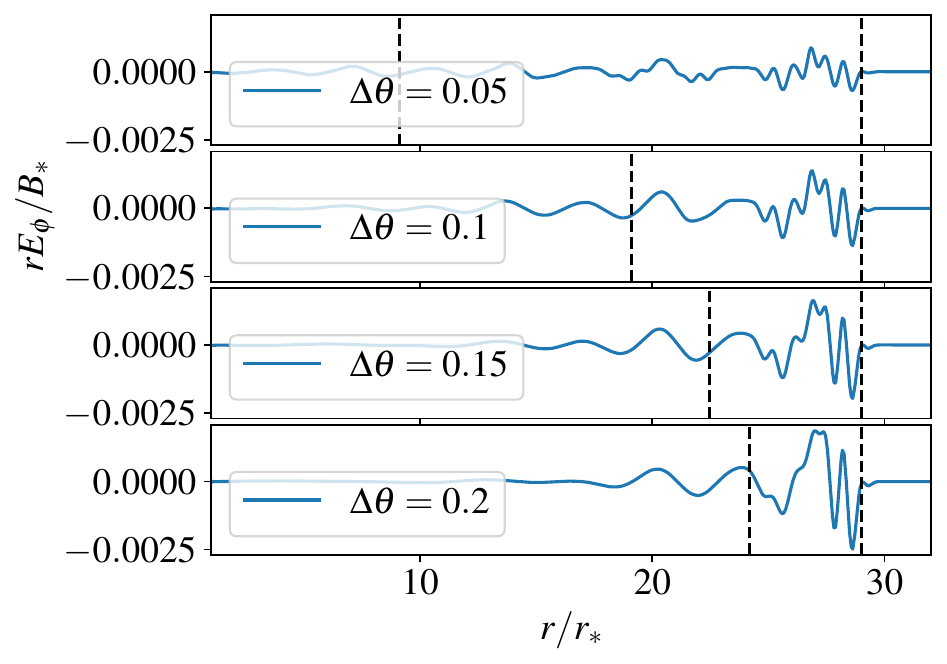}
    \caption{Cross sections of the fast wave train structure taken along $\theta = 3\pi/4$. From top to bottom the angular width of each star quake is $\Delta\theta = 0.05$, $\Delta\theta = 0.1$, $\Delta\theta = 0.15$, and $\Delta\theta = 0.2$. The black dashed lines indicate the time for the wave train to stretch to $2\ell$. All star quakes are initialized with $\theta_c = 0.8$, $d\omega_0 = 0.139c/r_*$, $cT = \ell$, and $\lambda_A = \ell/2$.}
    \label{Ap_fig:Structure_Low}
\end{figure}

In this section, we investigate the dependence of fast wave production on the angular width of the original \alfven{} wave. Consider an \alfven{} wave with train length $cT<2\ell$. In this regime, the fast waves produced are dominated by the low frequency component. We preform a series of simulations in which we vary the angular width of the flux tube centered around the field line originating from $\theta_c = 0.8$, while keeping the train length, wavelength, and amplitude of the \alfven{} wave constant. For this choice of $\theta_c$, $\ell \approx 3.3$. Figure \ref{Ap_fig:Structure_Low} shows a cross section of the fast waves taken along a $\theta = 3\pi/4$ for several different values of $\Delta\theta$. The low frequency component of the front of the wave is independent of the angular width, however the high frequency component is suppressed from significant dephasing when the angular width is large. We suspect that the low frequency component is unaffected by dephasing and the suppression of the tail is due to a stretching of the wave train. To quantify this, we compute the arc-length difference of the inner edge and outer edge of the flux tube ($\Delta\ell$) for each angular width ($\Delta \theta$). In section \ref{subsec:Structure} we argued that the constant component of $j_\phi$ carried by the \alfven{} wave will source a low frequency fast wave so long as the \alfven{} wave does not fill the entire flux tube. After one reflection the wave train is stretched by $\Delta\ell$. Once the \alfven{} wave train is stretched to $2\ell$, low frequency fast wave production ends. From this we approximate the train length of the fast waves as
\begin{equation}
    ct = \ell(2\ell - cT)/(\Delta\ell)
    \label{eq:fast_train_length}
\end{equation}
where $\Delta\ell$ depends on $\theta_c$ and $\Delta\theta$. In Figure \ref{Ap_fig:Structure_Low}, the distance between the two black dashed lines shows the train length computed using equation \ref{eq:fast_train_length}. We believe we are underestimating the duration of the fast wave train because the amplitude of the \alfven{} wave falls off near the edge of the flux tube. This makes the effective angular width of the flux tube smaller, resulting in a longer train length.

\begin{figure}[ht]
    \includegraphics[width=0.47\textwidth]{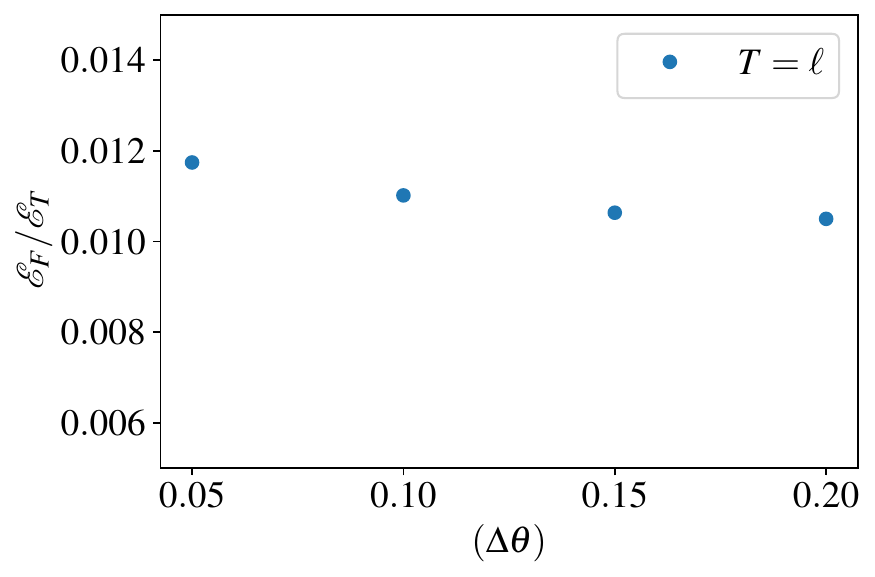}
    \caption{The conversion efficiency plotted against the angular width of the \alfven{} wave launched by the star quake, in the short \alfven{} wave train regime. The star quakes are all initialized with $\theta_c = 0.8$, $d\omega_0 = 0.139c/r_*$, $cT = \ell$, and $\lambda_A = \ell/2$.}
    \label{Ap_fig:Efficiency_Low}
\end{figure}

Figure \ref{Ap_fig:Efficiency_Low} shows the dependence of the conversion efficiency on the angular width. Although the fast wave train length decreases, the total energy converted to fast waves has limited dependence on the angular width because the front of the wave train is responsible for the most energetic fast waves. Because of the insignificant dependence on angular width, the low frequency portion of Equation \ref{eq:conversion_efficiency} has no dependence on angular width.

A comment is in place regarding the reduction of fast wave production when we introduce an initial phase shear in the \alfven{} wave by adding a time delay that depends on the launch angle in section \ref{subsec:long_term}, as shown in Figure \ref{fig:Shear}. However, several of our results indicate that the low frequency fast wave is less sensitive to dephasing. We see that the low frequency fast wave persists for much longer than the high frequency fast wave in Figure \ref{fig:SlowStructure}. We also see that the angular width, which affects how quickly the wave dephases, has little affect on the conversion efficiency in Figure \ref{Ap_fig:Efficiency_Low}. To understand this discrepancy, we argue that the drop in the conversion efficiency when shearing the initial \alfven{} wave occurs because shearing the \alfven{} wave train stretches the initial wave packet. This effectively initializes the wave with a train length as $cT = \ell + c\xi$. This reduction in low frequency fast wave production is also seen in Figure \ref{fig:Ratio}. Although we are unable to quantify the exact effect of dephasing on the conversion, we believe that the drop in conversion efficiency seen in Figure \ref{fig:Shear} is due more to the stretching of the \alfven{} wave train and less due to the phase mixing that this induces.

\begin{figure}[ht]
    \includegraphics[width=0.47\textwidth]{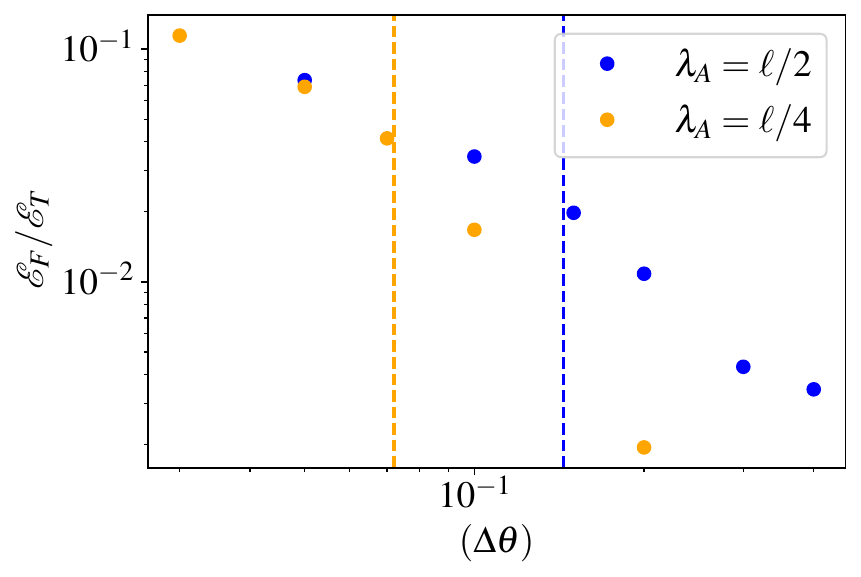}
    \caption{The conversion efficiency plotted against the angular width of the \alfven{} wave launched by the star quake, in the long \alfven{} wave train regime. Blue and orange correspond to star quakes with $\lambda_A = 0.5\ell$ and $\lambda_A = 0.25\ell$ respectively. The dashed lines show where the dephasing time is equal to $3\ell/2$. All other parameters of the star quakes are fixed with $\theta_c = 0.8$, $d\omega_0 = 0.139c/r_*$, and $cT = 2.5\ell$.}
    \label{Ap_fig:Efficiency_H}
\end{figure}

\begin{figure}[ht]
    \includegraphics[width=0.47\textwidth]{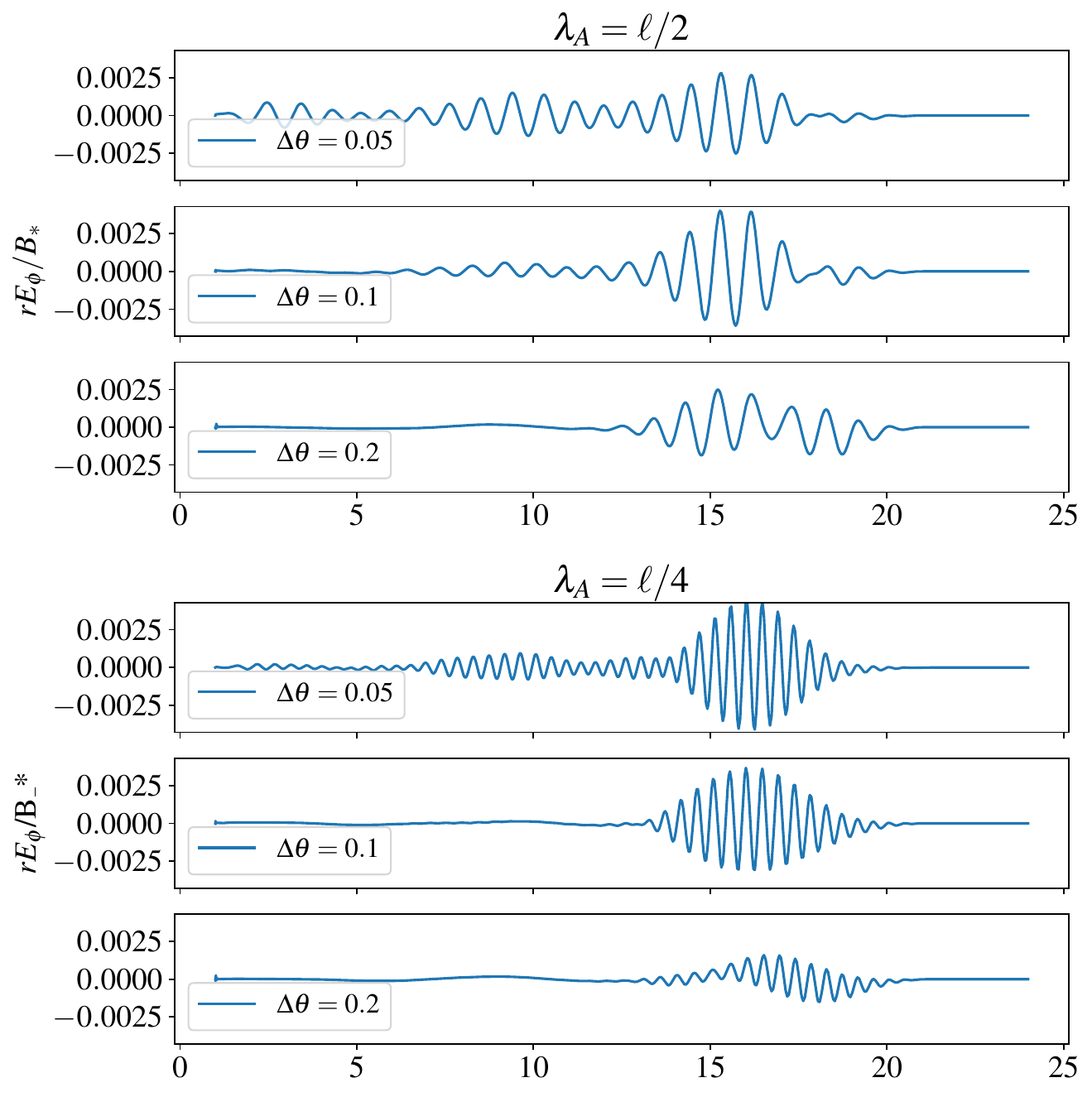}
    \caption{Cross sections of the fast wave train structure taken along $\theta = 3\pi/4$. The top three panels have $\lambda_A = 0.5\ell$ and the bottom three panels have $\lambda_A = 0.25\ell$. From top to bottom the angular width of each star quake is $\Delta\theta = 0.05$, $\Delta\theta = 0.1$, and $\Delta\theta = 0.2$, and again for the bottom three panels. All star quakes are initialized with $\theta_c = 0.8$, $d\omega_0 = 0.139c/r_*$, and $cT = 2.5\ell$.}
    \label{Ap_fig:Structure_H}
\end{figure}

We also explore the relationship between the angular width and the high frequency fast waves generated when $cT>2\ell$. Again using $\theta_c = 0.8$ as the central field line, we do a series of simulations that vary the angular width for two different fixed wavelengths. For all simulations, $d\omega_0 = 0.139$ and $cT = 2.5\ell$ are fixed. Figure \ref{Ap_fig:Efficiency_H} and Figure \ref{Ap_fig:Structure_H} show the conversion efficiency and the structure of the fast waves generated respectively. We see an overall drop in the high frequency fast wave production when the angular width of the \alfven{} wave increases. The relationship between the conversion efficiency and the angular width is complicated and does not have a well defined scaling. In section \ref{subsec:ConvEff}, we argue that the enhanced conversion seen in the long train length case is due to interacting \alfven{} waves. To explain the dependence of conversion efficiency on the \alfven{} wave angular width, we define a dephasing time to be the time it takes for the inside edge of the front of the \alfven{} wave to be completely out of phase with the outer edge of the front of the \alfven{} wave. When the desphasing time is shorter than the bouncing time on the flux tube ($t_{dephase} < \ell/c$), then the \alfven{} wave conversion will not be amplified by the interaction, causing the conversion efficiency to rapidly drop for large angular widths. The dashed lines in Figure \ref{Ap_fig:Efficiency_H} indicate the transition to fast wave production that dephases before the waves can self interact, amplifying the conversion. We calculate this using equation \ref{eq:angular_limit} as $\Delta\theta = \lambda_A \sin^3\theta_c/(6r_*\cos\theta_c)$. We emphasize that the necessary angular width for self interacting \alfven{} waves to amplify fast wave production goes to $0$ as $\sin^3\theta_c$ which drops quickly. That is, even for small angular widths, \alfven{} waves will dephase rapidly for high latitude star quakes.

Figure \ref{Ap_fig:Structure_H} shows a cross section of the fast waves taken along $\theta = 3\pi/4$ for the case of two different wavelengths. The \alfven{} wavelength for the top three panels is $\lambda_A = \ell/2$ and for the bottom three panels is $\lambda_A = \ell/4$. For $\lambda_A = \ell/2$ and $\lambda_A = \ell/4$, $t_{dephase} = \ell/c$ when $\Delta\theta \approx 0.14$ and $\Delta\theta \approx 0.07$ respectively. In section \ref{sec:discussion}, we approximated the train length in the long \alfven{} wave train regime as being equal to the \alfven{} wave train length. Figure \ref{Ap_fig:Structure_H} shows that simulations where $t_{dephase}<\ell/c$ have longer train lengths than the \alfven{} train lengths, however we still see the most significant conversion in a length approximately equal to the \alfven{} train length.

\bibliography{reference}{}
\bibliographystyle{aasjournal}

\end{document}